# Methods for extending viewing-angle of holographic image by using digital hologram with high numerical aperture

BYUNG GYU CHAE*

*Biomedical Imaging Group, Electronics and Telecommunications Research Institute, 218 Gajeong-ro, Yuseong-gu, Daejeon 305-700, South Korea*
*bgchae@etri.re.kr*

**Abstract:** We investigate the angular field of view (AFOV) of a holographic image reconstructed from the digital Fresnel hologram in holographic display. The theoretical analysis reveals that the AFOV of a holographic image is fundamentally determined by the hologram numerical aperture (HNA) other than a diffraction angle of pixel pitch of a pixelated modulator. This property is proved for various types of the digital holograms by using a numerical simulation and optical experiments. The high-HNA hologram reconstructs the image with a high viewing-angle, although the image contraction is inevitable due to the Nyquist sampling criterion. We propose the method for extending the viewing-angle of a holographic image in the manner of increasing the object size during the high-HNA hologram synthesis and removing the high-order aliasing images.

## 1. Introduction

The digital hologram in holographic displays is bandlimited by the finite pixel pitch of a digitally pixelated modulator [1-3]. The holographic image is reconstructed by illuminating the digital hologram with a coherent plane wave of wavelength $\lambda$. The space-bandwidth product (SBP) of the digital hologram is known to be a measure of its capacity for the reconstructed image [4,5]. The hologram function has an SBP corresponding to data capacity, which can be expressed using a lateral size $L$ and diffraction angle $\theta$ in the $(\xi, \eta)$ plane:

$$\text{SBP} = L_\xi L_\eta \frac{\theta_\xi \theta_\eta}{\lambda^2}. \tag{1}$$

The angle value $\theta$ is inversely proportional to a pixel pitch, and thus, a smaller pixel pitch leads to a wider diffraction field. For this reason, the pixel size $\Delta p$ has been known to determine the viewing-angle of a holographic image [6-9]:

$$\theta = 2\sin^{-1}\frac{\lambda}{2\Delta p}. \tag{2}$$

This interpretation makes it difficult for us to develop a commercial holographic display. Since the pixel size of a present spatial light modulator (SLM) is still on the scale of several micrometers, only a viewing-angle of several degrees can be obtained [10,11]. We also know from Eq. (1) that the size and viewing-angle have a trade-off relation for a constant hologram capacity. This smaller viewing-angle problem should be resolved to realize the holographic display. Most researches to settle the limitation of angular field of view (AFOV) have been carried out by expanding a diffraction zone with spatial and temporal multiplexing of the SLM [2-4,6-9], where the enormous data capacity is required to display a 3D scene even in one frame. Therefore, it is desirable to secure sufficient AFOV of a holographic image by using a commercial modulator without its multiplexing. The related researches still focus on enhancing a diffraction angle [12,13]. To overcome this limitation, the deeper analysis to identify the cause of viewing-angle change must take precedence.

The diffracted Fresnel field is well expressed as the convolution of object field and impulse response function $h(x, y; z)$ [1,14]:

$$h(x, y; z) = \frac{e^{ikz}}{i\lambda z} \exp\left[i\frac{\pi}{\lambda z}(x^2 + y^2)\right], \quad (3)$$

where $k$ is wavenumber of $\lambda/2\pi$ and $z$ is a propagation distance. To avoid an aliasing effect in the digital hologram synthesis, the sampling rate of object field at respective distances is restricted by the maximum spatial frequency $f_{x,\max}$ in $x$-coordinate, which can be obtained with a quadratic phase $\phi(x, y)$ of Eq. (3):

$$f_{x,\max} = \frac{1}{2\pi}\left|\frac{\partial \phi(x, y)}{\partial x}\right|_{\max} = \frac{x_{\max}}{\lambda z}. \quad (4)$$

The maximum spatial frequency depends on both a distance and object field size. The sampling rate $f_s$ of object field should satisfy the Nyquist sampling criterion [15,16]:

$$f_s \geq 2 f_{x,\max} = \frac{2 x_{\max}}{\lambda z}. \quad (5)$$

Considering the sampling rate digitizing the object, the size of object field should be reduced with decreasing a synthesis distance to avoid an aliasing error. As shown in Fig. 1, when the digital holograms encoded in the SLM have a constant magnitude, the object size will be adjusted according to a synthesis distance [17-19]. The discrete Fresnel transform used as a single Fourier transform is properly operated in accordance with this condition as long as the object size is not much larger than the hologram size. In holographic display, the reconstruction process is a backpropagation from the digital hologram to the image plane. In Fig. 1, since the digital holograms synthesized at various distances have the same pixel size, the viewing-angle of various reconstructed images seems to be invariant based on Eq. (2). However, up to now, there has been no detailed study to elucidate whether or not the viewing-angle is constant irrespective of a geometry.

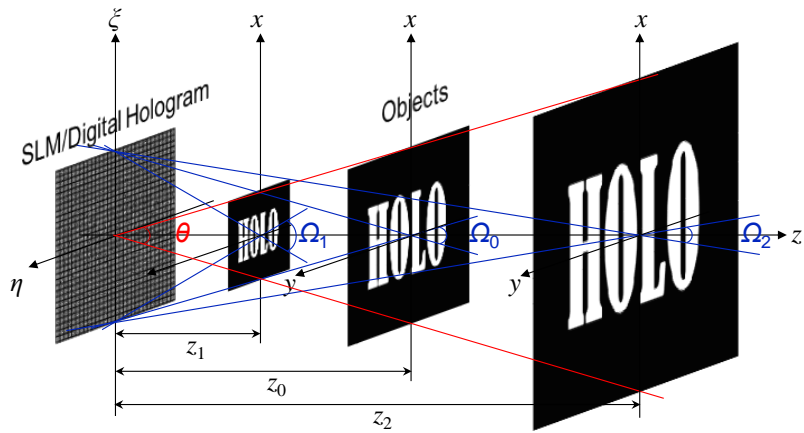

Fig. 1. Configuration of digital hologram and objects during the hologram synthesis. The object size decreases at lower distance from the digital hologram, based on the Nyquist criterion, and the angles from the blue lines and red lines mean the values for the numerical aperture of the digital hologram and diffraction zone due to a pixel pitch, respectively.

In this study, we carry out the analysis of the AFOV of a reconstructed image for the sampled hologram on a pixelated modulator, and explain the AFOV dependent on the numerical aperture of the hologram other than the pixel pitch. We perform the numerical simulation investigating the change in the viewing-angle of images for various types of digital holograms in the Fresnel diffraction regime. The diffraction fringes propagated from the reconstructed image are simulated, which allows us to evaluate its viewing-angle by measuring the increment of an active diffraction field. Optical experiment is also conducted to confirm our analysis. We apply this analysis to search for the method extending the viewing-angle of a reconstructed image.

## 2. AFOV of holographic image dependent on hologram numerical aperture of digital hologram

*2.1 Diffraction properties for sampled hologram on pixelated modulator*

Let us consider the sampled hologram $g_s(\xi,\eta)$ on the pixelated modulator with rectangular pixels of the pixel interval $p$ and pixel size $\Delta p$,

$$g_s(\xi,\eta) = \sum_{n_\xi=-\infty}^{\infty} \sum_{n_\eta=-\infty}^{\infty} \left[ g(n_\xi p_\xi, n_\eta p_\eta) \mathrm{rect}\left( \frac{\xi - n_\xi p_\xi}{\Delta p_\xi}, \frac{\eta - n_\eta p_\eta}{\Delta p_\eta} \right) \right], \quad (6)$$

where rect() is a rectangular function. In the Fresnel diffraction regime, the diffractive object wave propagating from the hologram can be represented as a convolution form of two terms [12]:

$$O(x,y) = \frac{ie^{-ikz}}{\lambda z} e^{-i\frac{k}{2z}(x^2+y^2)} \iint g_s(\xi,\eta) \exp\left[ i\frac{2\pi}{\lambda z}(x\xi + y\eta) \right] d\xi d\eta$$
$$* \iint \exp\left[ -i\frac{\pi}{\lambda z}(\xi^2 + \eta^2) \right] \exp\left[ i\frac{2\pi}{\lambda z}(x\xi + y\eta) \right] d\xi d\eta, \quad (7)$$

where the normally incident plane wave with unit amplitude is used. The integral form of the first line represents the Fourier spectrum of the sampled hologram:

$$\iint g_s(\xi,\eta) \exp\left[ i\frac{2\pi}{\lambda z}(x\xi + y\eta) \right] d\xi d\eta = \Delta p_\xi \Delta p_\eta \, \mathrm{sinc}\left( \frac{\pi x \Delta p_\xi}{\lambda z} \right) \mathrm{sinc}\left( \frac{\pi y \Delta p_\eta}{\lambda z} \right)$$
$$\times \sum_{n_\xi=-\infty}^{\infty} \sum_{n_\eta=-\infty}^{\infty} \left\{ g(n_\xi p_\xi, n_\eta p_\eta) \exp\left[ i\frac{2\pi}{\lambda z}(n_\xi p_\xi x + n_\eta p_\eta y) \right] \right\}. \quad (8)$$

This equation describes the modulation of the periodic Fourier spectrum by the envelope of a sinc function along the *x*- and *y*-axis. The summation term indicates the periodic Fourier spectrum through the Poisson summation formula,

$$\sum_{\alpha=-\infty}^{\infty} \sum_{\beta=-\infty}^{\infty} G\left[ \frac{1}{\lambda z}\left( x - \frac{\lambda z}{p_\xi}\alpha \right), \frac{1}{\lambda z}\left( y - \frac{\lambda z}{p_\eta}\beta \right) \right]. \quad (9)$$

By using the diffraction relations, $p_\xi \sin\theta_\alpha = \alpha\lambda$ and $p_\eta \sin\theta_\beta = \beta\lambda$, the convolution expression of Eq. (7) yields

$$O(x,y) = \frac{ie^{-ikz}}{\lambda z} e^{-i\frac{k}{2z}(x^2+y^2)} \Delta p_\xi \Delta p_\eta \, \mathrm{sinc}\left( \frac{\pi x \Delta p_\xi}{\lambda z} \right) \mathrm{sinc}\left( \frac{\pi y \Delta p_\eta}{\lambda z} \right) \left\{ \sum_{\alpha=-\infty}^{\infty} \sum_{\beta=-\infty}^{\infty} \right.$$

$$\times \iint g(\xi,\eta) \exp[ik(\xi \sin\theta_\alpha + \eta \sin\theta_\beta)] \exp\left[-i\frac{\pi}{\lambda z}(\xi^2+\eta^2)\right] \exp\left[i\frac{2\pi}{\lambda z}(x\xi+y\eta)\right]d\xi d\eta \Bigg\}. \quad (10)$$

We find that the high-order diffraction beams are generated from the sampled hologram on a pixelated modulator. Figure 2 illustrates a schematic diagram for the generation of high-order images in one-dimensional space. The Fraunhofer diffraction patterns are formed at a distance close to the hologram aperture. In the present modulator with a pixel size of about 8 μm, the far-field region appears at a submillimeter distance [1]. When $z$ is a synthesis distance of the digital hologram, replica images will be displayed within a lateral space at the interval of $\lambda z/p$ [12,20,21]. The images are modulated by the envelope of a sinc function, where the modulated pattern is decided by the ratio of pixel size and pixel pitch called as a fill factor, and the position of images can be arbitrarily controlled in terms of a phase shift [12]. This behavior is well observed in optical experiments. In holographic display, only 1$^{st}$-order image is adopted, and thus, the maximum viewing-angle seems to be the 1$^{st}$-order diffraction angle. However, the viewing-angle $\Omega$ of a reconstructed image does not need to be equal to 1$^{st}$-order diffraction angle $\theta$, because the restored image in 1$^{st}$-order diffraction zone is affected from whole pixels of a pixelated modulator, as shown in Fig. 2. That is to say, the spherical wavelets in all pixels contribute to the image formation as well as a modulating sinc function.

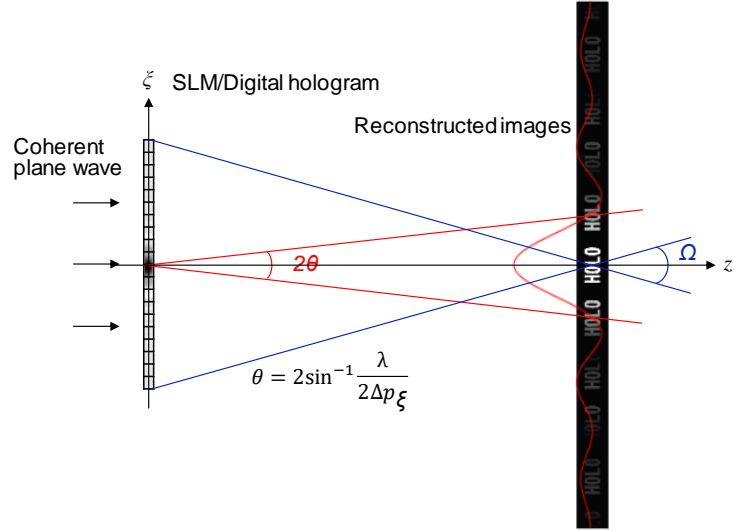

Fig. 2. Schematic diagram for the generation of high-order images from the sampled digital hologram. For convenience, the diagram is drawn for the complex hologram in one-dimensional space.

We note that the pixelated structure contributes only to the formation of a periodic diffraction zone. In other words, we may interpret most of the specification of a holographic image separately from the pixel structure. The integral of the respective diffraction zones in Eq. (10) looks like a diffraction formula for continuous signals, where the near-field region covers a relatively long distance according to a hologram aperture size. From this conjecture, we know that the AFOV of a holographic image will not simply depend on the diffraction extent of pixel pitch of a spatial modulator. The pixel pitch just causes the diffraction zone of Eq. (2). Here, the object field size seems to be limited to the diffraction area due to the Nyquist criterion, however even the object field beyond the diffraction area of a pixel pitch can be calculated by increasing an object resolution, as confirmed in Section 5.

## 2.2 Dependence of viewing-angle on the resolution limit of holographic image

The integral term of the respective diffraction zones in Eq. (10) is the convolution of the hologram field and inverse impulse response function $h(x, y; -z)$. We treat this integral as a diffraction propagation from the analog hologram with a finite aperture size $L$:

$$O(x,y) = g(x,y)\text{rect}\left(\frac{x}{L}, \frac{y}{L}\right) * h(x, y; -z), \tag{11}$$

where the lowest order term, $\theta = 0$ is considered. Above equation is expanded as follows,

$$O(x,y) = \frac{e^{ikz}}{i\lambda z}\exp\left[-\frac{i\pi}{\lambda z}(x^2 + y^2)\right]FT\{g(\xi,\eta)\} * FT\{h(x,y;-z)\} * FT\left\{\text{rect}\left(\frac{\xi}{L}, \frac{\eta}{L}\right)\right\}. \tag{12}$$

We find that the term of Fourier transform **FT** of a rectangular function plays a role in the convolutional kernel for an image resolution. To explore an intrinsic resolution of restored image, we adopt a hologram function only for a point object $\delta(x, y)$, since the real object with a finite extent can be regarded as a collection of individual point objects. The complex hologram for a point object is equal to the impulse function of Eq. (3), and thus, the object image is calculated to be in the form of a sinc function:

$$O(x,y) = \left(\frac{\pi L}{\lambda z}\right)^2 \exp\left[-\frac{i\pi}{\lambda z}(x^2 + y^2)\right]\text{sinc}\left(\frac{\pi x L}{\lambda z}\right)\text{sinc}\left(\frac{\pi y L}{\lambda z}\right). \tag{13}$$

In case of an infinite aperture size, the restored image appears to be a delta function. The width of the first maximum peak of a sinc function indicates an image resolution limit resolving the closest points irrespective of a phase curvature in front of a sinc function [20,22,23], based on the Abbe criterion. We consider a digital hologram having pixels of $\Delta\xi$ and $\Delta\eta$ with a 100% fill factor. Since an aperture size is $N_\xi \Delta\xi \times N_\eta \Delta\eta$, the resolution limits $R_{x,y}$ become

$$R_x = \frac{\lambda z}{N_\xi \Delta\xi} \text{ and } R_y = \frac{\lambda z}{N_\eta \Delta\eta}. \tag{14}$$

These values are explained by the hologram numerical aperture (HNA). If we assume that the specification of pixels is the same for both axes, the HNA in a free space is geometrically given by

$$\text{HNA} = \sin \Omega_{\text{HNA}} = \frac{N_\xi \Delta\xi}{2z}. \tag{15}$$

Thus, the resolution limit $R_{\text{Abbe}}$ of the Abbe criterion in a hologram imaging procedure is expressed as [14,24]

$$R_{\text{Abbe}} = \frac{\lambda}{2\text{HNA}}. \tag{16}$$

In a point-like object, the converging spherical wave from the digital hologram forms an object image, and one imagines that the diverging wave from the object image propagates to a free space. The converging and diverging spherical waves have a mirror symmetry with respect to an imaging plane. The viewing-angle of the point-like object image should be directly related to the HNA, which could become a fundamental criterion for the AFOV of holographic images, as depicted in Fig. 2. The hologram acquisition and its image reconstruction is a coherent imaging process [24,25]. The object field information is acquired through an optical lens in a conventional imaging process, whereas in the holography the field information is directly recorded on the digital hologram and the object is reconstructed in the image plane numerically or optically. During its reconstruction process, the intrinsic resolution of object field will be obtained. Finally, the viewing-angle $\Omega$ of a reconstructed image can be written in the form:

$$\Omega = 2\sin^{-1}\left(\frac{\lambda}{2R_{\text{Abbe}}}\right). \qquad (17)$$

We find that the viewing-angle is fundamentally determined by a resolution limit of a holographic image rather than a pixel pitch of digital hologram like Eq. (2).

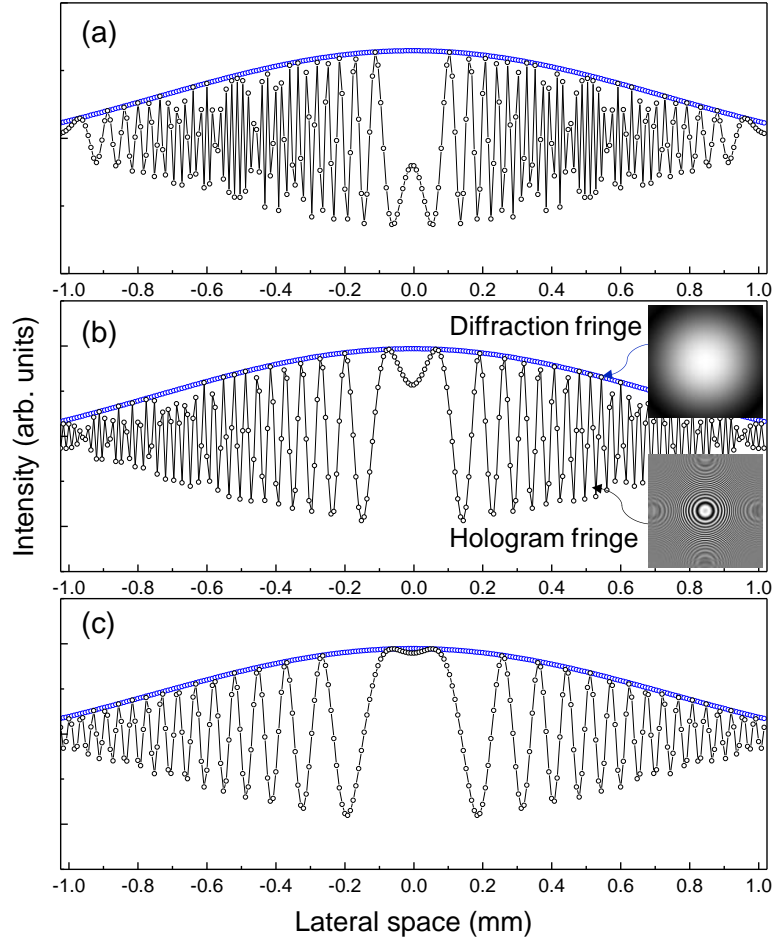

Fig. 3. Typical example of hologram fringes for point-like objects located at a different distance. The data is calculated with circular objects of (a) 4-μm size located at a 15.4-mm distance, (b) 8-μm size located at a 30.8-mm distance, and (b) 16-μm size located at a 61.6-mm distance, based on the Riemann integral in the Rayleigh-Sommerfeld diffraction formula.

## 2.3 Analysis of digital hologram fringe

The Fresnel diffraction from object with a finite extent can be analyzed by the following equation:

$$g(\xi,\eta) = \frac{e^{ikz}}{i\lambda z} \exp\left[\frac{i\pi}{\lambda z}(\xi^2 + \eta^2)\right] FT\left\{O(x,y)\exp\left[\frac{i\pi}{\lambda z}(x^2 + y^2)\right]\right\}. \quad (18)$$

In Section 1, we investigate the aliasing phenomenon of digital hologram arising only from the sampling of object field where the quadratic phase term in parenthesis of **FT** is treated. For simplicity, we adopt one-dimensional interpretation hereafter. Based on the Fourier analysis of the discrete Fresnel transform, the pixel resolution $\Delta\xi$ of hologram field is defined in accordance with resolution $\Delta x$ of the object field [25]:

$$\Delta\xi = \frac{\lambda z}{N_x \Delta x}. \quad (19)$$

As previously described, the Fresnel field in the hologram plane is well calculated under sampling condition:

$$z \geq \frac{N_x \Delta x^2}{\lambda}. \quad (20)$$

However, although the digital hologram synthesized at a closer distance such as $z_1$ in Fig. 1 satisfies above condition, the undersampling of diffraction field takes place because of a rapid oscillation of a quadratic phase factor in front of **FT** operation. To elucidate this character apparently, the analysis of aliasing error from the aspect of hologram plane should be performed by using a Fresnel factor in the $(\xi,\eta)$ coordinates [15,16]. According to this approach, the well-sampling operation is possible only at a constant distane, $z = N_x \Delta x^2/\lambda$. The aliasing error of hologram fringe could be inevitable from the undersampling of oscillating phase of a Fresnel prefactor when the sampling pitch of hologram field is lower than that of object field. In this case, the reconstruction process of image will be obstructed by the aliasing error, and furthermore, the sampling condition in the reconstruction process violates the Nyquist criterion. However, we will show the robust retrieval of original image despite of this type of aliasing error of hologram fringe. The aliasing error generated from the hologram synthesis can be compensated in the reconstruction process. The detailed analysis is out of present research scope, which will be clarified in the subsequent research [26].

Figure 3 is a typical example of the hologram fringes for point-like objects located at a different distance. For convenience, we assume that the object size is put to be its image resolution. The point-like object generates the spherical wave diverging radially, where the real or imaginary hologram has a concentric fringe similar to the sinusoidal Fresnel zone plate [24,27]. The dense hologram fringe will be synthesized at a close distance, and as illustrated in Fig. 3(a), at much closer distance aliasing fringe arises from the undersampling of oscillating phase. We know that only a spatial information of the point-like object is encoded in the fringe shape.

We note that the intensity profile of diffractive wave is resulted from the **FT** operation of the product of object field and quadratic phase term. This intensity profile called as Airy pattern determines the image resolution. The $\text{SBP}_{1d}$ of synthesized hologram at the $z$-distance can be interpreted in two components related to the numerical aperture and hologram fringe frequency:

$$\text{SBP}_{1d} = N_\xi \Delta\xi \cdot \frac{1}{\Delta\xi}. \quad (21)$$

The hologram size $N_\xi \Delta\xi$ presents the propagating window size of the Fourier spectrum with respect to the object resolution $\Delta x$, which defines the HNA:

$$N_\xi \Delta\xi = \frac{\lambda z}{\Delta x} = 2z\sin(\Omega_{\text{HNA}}). \quad (22)$$

The 1st order diffraction fringe from object field will be sufficient for the hologram, whose size means the magnitude of hologram aperture. Meanwhile, the hologram fringe frequency correlates with oscillating phase of a preceding Fresnel factor, which points out only the diffraction extent with respect to hologram pixel pitch:

$$\frac{1}{\Delta \xi} = \frac{\sin \theta}{\lambda}. \tag{23}$$

It is interesting that in digital hologram, the fundamental origin in defining the resolution of restored image is the HNA other than the maximum spatial frequency of hologram fringe. In spectrum values of **FT** operation in Eq. (18), the sampling is complete even at a close distance because the sampling condition is well satisfied by Eq. (20). In this situation, only an oscillating phase before **FT** spectrum is undersampled. In an analog hologram acquisition, a hologram fringe with a high spatial frequency can be obtainable up to the resolution limit of high sensitive photographic film, where the numerical aperture will be directly represented as the maximum spatial frequency of hologram fringe, $\lambda f_{\max}$ because a diffraction spot of object is regarded as a delta function.

*2.4 Viewing-angle variation in terms of hologram numerical aperture*

In holographic display, the object image is optically focused on the image plane, and thus the ray of the object image field produces the viewing-angle with the same size as a double angle of the HNA. As previously described in the discrete Fresnel transform, the relationship of pixel resolution of object and hologram field reveals the dependence of pixel resolutions $\Delta x$ of the object image field upon the HNA by itself:

$$\Delta x = \frac{\lambda z}{N_\xi \Delta \xi} = \frac{\lambda}{2 \sin(\Omega_{\mathrm{HNA}})}. \tag{24}$$

Since the resolution limit $R_{Abbe}$ coincides with the pixel size $\Delta x$ of object image in this case, the viewing-angle $\Omega$ of the holographic image is written by

$$\Omega = 2 \sin^{-1} \left( \frac{N_\xi \Delta \xi}{2z} \right), \tag{25}$$

where $z$ becomes the imaging or synthesis distance. As shown in Fig. 4(a), if we use a SLM with same pixel pitch to load the hologram fringe, the viewing-angle decreases with increasing the synthesis distance, and the image resolution gets worse as well. Here, the lateral size of the digital hologram becomes an aperture size.

On the other hand, in the digital hologram synthesized with maintaining the constant resolution of object, the viewing-angle will not change depending on a synthesis distance, in Fig 4(b). Particularly, at a distance far away from the object the hologram captures a part of its diffractive wave, while at a close distance, the whole diffraction fringe does not cover the total area of a SLM. This behavior acts as a low-pass filtering for hologram fringe to satisfy the Nyquist criterion. Here, the resolution limit $R_{Abbe}$ having the pixel size $\Delta x$ is constant, $\Delta x = \Delta \xi$. The viewing-angle $\Omega$ is given by

$$\Omega = 2 \sin^{-1} \left( \frac{\lambda}{2 \Delta \xi} \right). \tag{26}$$

Even though the whole diffractive wave is not displayed on a SLM, the viewing-angle maintains, where the lateral size of digital hologram does not become an aperture size. As will be disclosed in the numerical simulation, the convolution method keeps a resolution during the synthesis and its reconstruction process, and thus the viewing-angle remains invariable. However, in an optical reconstruction process for this hologram synthesized at a far away from object, the image resolution will decrease because of the HNA reduction. The description of Figs. 4(a) and 4(b) are related to the properties for the holograms synthesized by the conventional Fresnel transform and convolutional method, respectively.

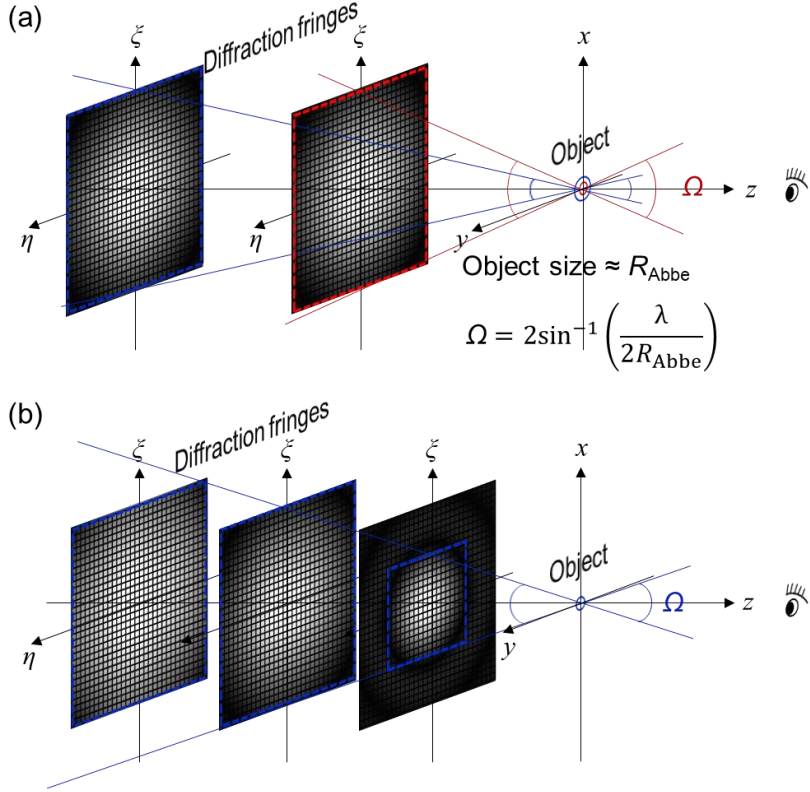

Fig. 4. Schematic diagram of analysis on AFOV of the holographic image for the digital hologram of the point-like object. The HNA for the digital hologram synthesized (a) by varying the object resolution and (b) with maintaining the constant resolution of the object.

## 3. Numerical analysis of AFOV of Fresnel holographic images

### 3.1 AFOV characteristics for holograms synthesized using a conventional Fresnel transform

For a numerical simulation, the diffractive wave field propagated from the object field is expressed as the discrete Fresnel transform [25]. The fields are digitized on rectangular raster with steps $\Delta\xi$ and $\Delta\eta$ in the output $(\xi,\eta)$ plane and $\Delta x$ and $\Delta y$ in the input $(x,y)$ plane:

$$g(m\Delta\xi, n\Delta\eta) = \frac{e^{ikz}}{i\lambda z}\exp\left\{i\frac{\pi}{\lambda z}\left[m^2\Delta\xi^2 + n^2\Delta\eta^2\right]\right\}$$
$$\times \boldsymbol{DFT}\left[O(s\Delta x, t\Delta y)\exp\left\{i\frac{\pi}{\lambda z}\left[s^2\Delta x^2 + t^2\Delta y^2\right]\right\}\right]. \qquad (27)$$

The second line equation indicates the discrete Fourier transform $\boldsymbol{DFT}$ of the product of the input field and a quadratic phase term.

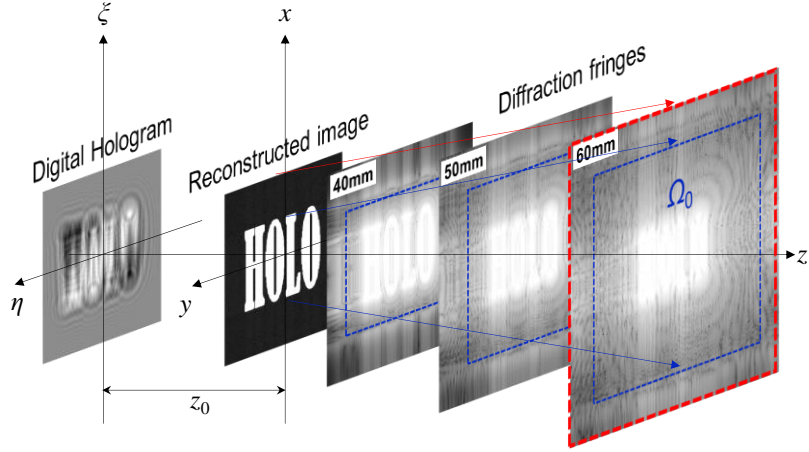

Fig. 5. Numerical studies for observing AFOV of the holographic image. The ratio of the lateral and longitudinal scale is arbitrarily resized for convenience. The red box and blue box indicate the total diffraction zone and active diffraction fringe, respectively.

Figure 5 illustrates the numerical studies for observing the AFOV of a holographic image. The digital hologram is synthesized by using the MATLAB code based on Eq. (27), and the real-valued hologram is displayed. The 'HOLO' letter object placed at $z_0$ distance in Fig. 1 is used, where $z_0$ is defined as

$$z_0 = \frac{N_x \Delta x^2}{\lambda}. \quad (28)$$

The object and hologram with 256×256 size have the same pixel pitch of 8 μm. The coherent plane wave has 532-nm wavelength, and in this condition, the distance $z_0$ is calculated to be 30.8 mm.

To investigate an angle of view of the reconstructed holographic image, the diffraction fringes far away from the imaging plane are numerically calculated via the reverse transform of Eq. (27), where the diffraction fringe is an intensity pattern of the diffracted wave. For convenience, a complex amplitude hologram is adopted. Considering a commercial pixelated modulator, the real-valued or imaginary-valued hologram will be realistic. In in-line holographic system, the overlap of a conjugate image makes it difficult to measure the diffraction fringe change. Therefore, the complex hologram is chosen for analyzing an angle of view of the holographic images hereafter.

The diffraction fringes are displayed with a logarithmic scale to mitigate the energy concentration at the origin in the Fourier space. As shown in Fig. 5, we can observe the apparent diffraction fringe corresponding to the letter image propagation, while it is not clearly distinguishable in a linear-scale image. This active area in an inset box increases with increasing a reconstruction distance. The strip patterns outside the active area of the diffraction fringes arise from an aliasing effect due to spatially limited signal. We can see this aliasing phenomenon even in the restored image in the image plane. The total field in the discrete Fresnel transform varies in linear proportion to a reconstruction distance. The pixel resolution $\Delta x$ of a diffraction field at an $x$-coordinate is determined from the relation given in Eq. (24). The increment ratio of the field sizes indicates the diffraction angle, which is another form of Eq. (2), assuming that the angle is small. The angle value relevant to the diffraction zone is estimated to be about 3.81°. The active diffraction area in the blue box of Fig. 5 changes from 960 μm at 30.8-mm distance of the image plane to 3304 μm at 60-mm distance. The viewing angle $\Omega_0$ of 4.59° is estimated from a growth rate of the diffraction fringe along a distance. This value is a little large when compared to the diffraction angle due to the pixel pitch, but it looks like that an observable view in terms of the active diffraction fringe gets enlarged in similar proportion to the total view.

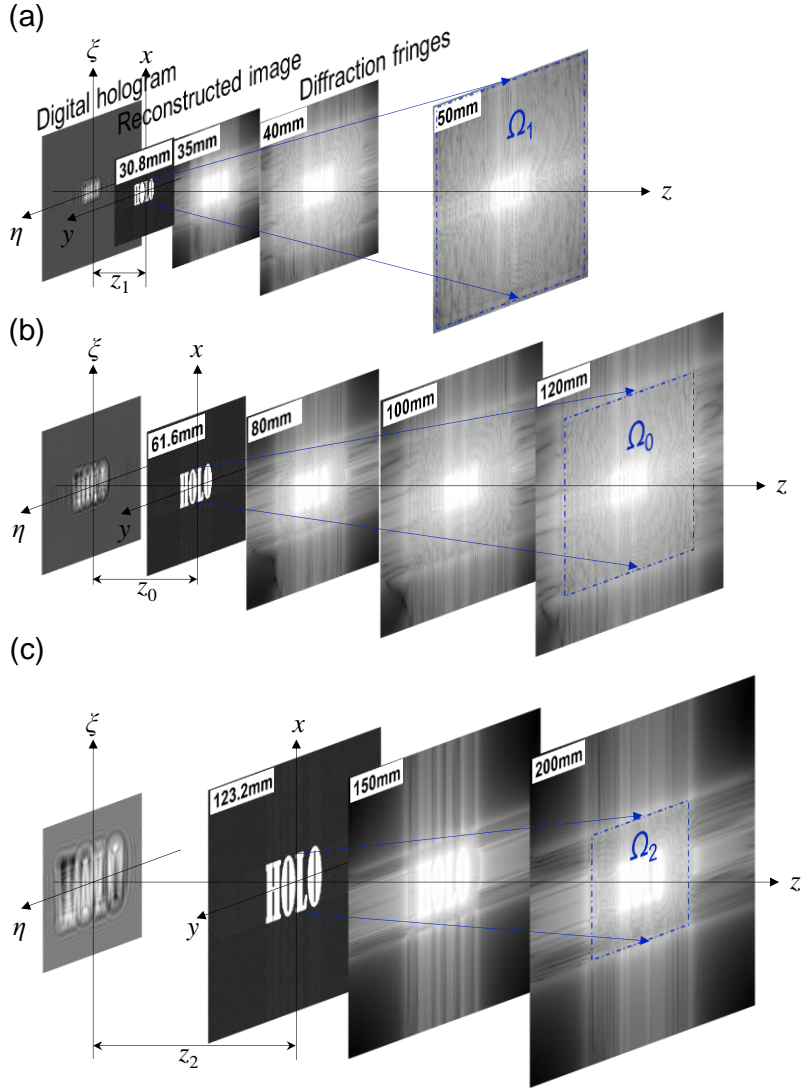

Fig. 6. Diffraction fringes propagated from the reconstructed images for the digital holograms synthesized at various distances of (a) 30.8 mm, (b) 61.6 mm, and (c) 123.2 mm.

Figure 6 is the simulation results for the digital holograms synthesized at various distances. Three kinds of digital holograms are prepared for the objects located at distances of a half and a doubling of $z_0$ as well as $z_0$ distance. To compare their viewing-angle variations clearly, all the objects are enlarged with 512×512 size using the zero-padding. The ratio of active area to an opaque background was confirmed to be irrelevant to the change in the viewing-angle. The small ratio of active area enables us to investigate the viewing-angle variation apparently. In general, since the holographic image is displayed on the opaque background in holographic display, this approach could be reasonable.

The pixel pitch of all the holograms is fixed to be 8 μm. The diffraction behavior of the hologram at a $z_1$-distance in Fig. 1 is displayed in Fig. 6(a). We calculate the $z_1$-distance of 30.8 mm, where the object pixel size is 4 μm and thus, the reconstructed image size is half of the hologram size. The increasing rate of the total field of the diffraction fringe away from the image plane is same as that in Fig. 5 because of the same 8-μm pixel size; however, the active diffraction region reveals a rapid increase, whose diffraction fringe occupies the whole area at 60-mm distance. The active area increases from 480 μm at 30.8-mm distance to 3039 μm at 50-mm distance. The viewing-angle $\Omega_1$ is calculated to be approximately 7.62°, whose value is approximately twice the diffraction angle of 8-μm pixel.

Figure 6(b) depicts the diffraction behavior of a reconstructed image for the hologram made at $z_0$-distance. The $z_0$-distance is 61.6 mm because the field size of the hologram and object is doubled with compared to that in Fig. 5. The viewing-angle $\Omega_0$ estimated from the increase of the active diffraction fringe is about 4.08°, which is a similar value in Fig. 5. The numerical results for the hologram made at $z_2$-distance are appeared in Fig. 6(c). The digital hologram is located at 123.2-mm distance from the object image with a pixel size of 16 μm, in Fig. 1. The active diffraction area increases modestly with a reconstruction distance as compared with previous results. The angle $\Omega_2$ is estimated to be 2.37°, which is close to half of the diffraction angle for 8 μm pixel size.

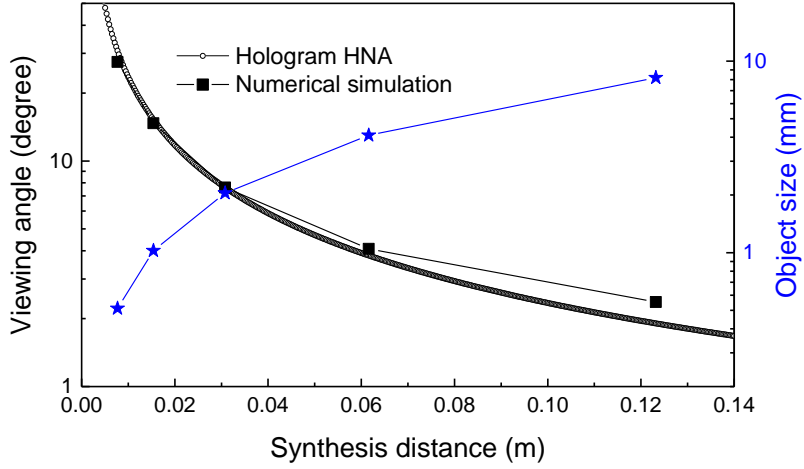

Fig. 7. Viewing-angle change in the reconstructed image for the digital holograms synthesized using a conventional Fresnel method. The object size variation is displayed.

The above results indicate that the AFOV of a reconstructed image cannot be simply determined by the diffraction angle of the pixel pitch of a spatial modulator. We find that as described in Section 2, the AFOV is rather decided in terms of the numerical aperture of the hologram. This is the case of the description in Fig. 4(a), where the viewing-angle $\Omega$ corresponds to Eq. (25). Figure 7 is the plot of the viewing-angle change in the reconstructed image as a function of a synthesis distance. The variation of the viewing-angle matches well with the angle $\Omega$ obtained from the HNA. The angle value reaches 27.5° at a 7.7-mm distance. Here, the upper bound of the angle will be limited within the Fresnel approximation, but this condition is known to be overly stringent [1]. Our analysis can be also extended to the Rayleigh-Sommerfeld region, and thus the higher angle can be obtainable in principle. The smaller synthesis distance results in a larger numerical aperture, which generates the reconstructed image with a wide viewing-angle, while it is inevitable the shrinkage of image size.

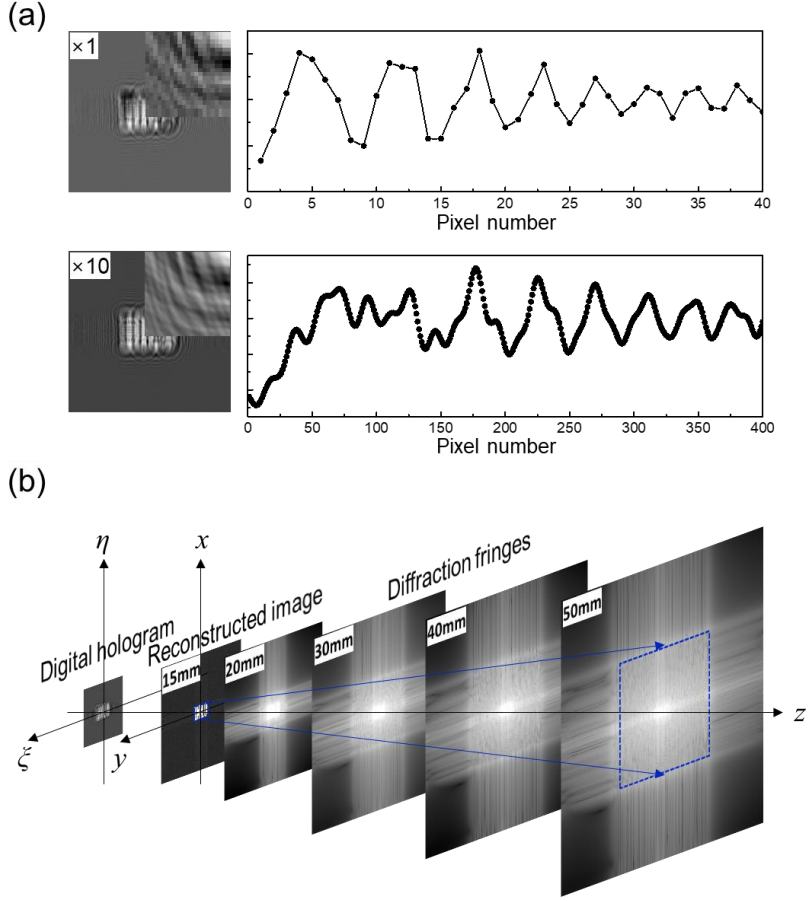

Fig. 8. Viewing-angle change in the holographic image through the upsampling process of the hologram. (a) Digital hologram fringes and pixel value distributions with respect to unenhanced resolution and ten-fold enhanced resolution. (b) Reconstructed diffraction fringes using the digital hologram with four-fold enhanced resolution.

*3.2 AFOV characteristics for upsampled hologram*

We apply this method of analyzing the field view to the upsampled hologram. In Eq. (19), a lower pixel value $\Delta\xi$ of the hologram can be obtained by varying the number of object plane pixels. When the number of pixels is $w$ times larger, the resolution of output field increases $w$ times,

$$\left(w \cdot N_\xi\right)\frac{\Delta\xi}{w}. \tag{29}$$

If the field size of the object is doubled in terms of the zero-padding while all other parameters are fixed, the pixel size of the digital hologram will be reduced to half. In a vector-matrix notation, the column vector in the hologram plane is expressed as the multiplication of the Fresnel matrix and object vector. The Fresnel matrix is composed of the Fourier kernel and a quadratic phase term in Eq. (27). The Fresnel matrix elements corresponding to the extended parts of object space are multiplied by the zero values of the object vector. This is an upsampling of the digital hologram, which is known as zero-padding technology [14]. This numerical hologram has higher SBP than the original. From Eq. (1), as the lateral size of the hologram is fixed, the higher SBP can be expected to increase the viewing-angle for the restored image.

Figure 8(a) shows the digital hologram fringes and pixel value distributions via the upsampling process. The digital hologram of 256×256 size with pixel pitch 8 μm is used. The pixel pitch of the letter object placed at a 15.4-mm distance is calculated to be 4 μm. The fringe upsampled with 10-times enhanced resolution has a finer pixel pitch of 0.8 μm. Each upsampled subpixel makes a smooth connection with the nearest neighbor pixels in the magnified fringe. The pixel graph in a section of fringe exhibits a wiggle not observed in the original fringe.

Simulation results for the viewing-angle change of holographic image through the upsampling process of a hologram fringe are illustrated in Fig. 8(b). The reconstructed diffraction fringes are about the digital hologram with the resolution enhanced four-fold. The reconstructed letter image at the same distance of 15.4 mm has a field size 4 times larger than the original field size of 1024 μm. The diffraction angle by the increment of total field is 15.2°, which is the angle value for four-fold enhanced resolution with a 2-μm pixel size. On the other hand, the active diffraction fringe from the letter image enlarges in smaller proportion to the total view. The estimated viewing-angle is approximately 8.4°. In this geometry, since the object size with the 4-μm pixel is a half of the hologram size, it generates two-fold increase in the viewing-angle based on the explanation in Fig. 7. Therefore, this value is rather close to the original angle for the pixel of 8 μm in the hologram with the unenhanced resolution. We find that the upsampled hologram does not affect the viewing-angle enlargement, but only enlarges the diffraction viewing-zone due to the pixel pitch. The object image resolution remains constant, and thus, the HNA does not change where the viewing-angle maintains.

*3.3 AFOV characteristics for holograms synthesized using a convolution method*

The digital hologram can be also synthesized by using a convolutional approach, where the pixel resolution of the input plane and output plane has the same value. In the convolution method, the diffractive wave filed is represented as the convolution of the input field and impulse response function [14]. The Fourier transform of the impulse response function is called as a spatial frequency transfer function,

$$H(u,v) = e^{ikz} \exp\left[-i\pi\lambda z(u^2 + v^2)\right]. \tag{30}$$

The discrete form of the output field is written using a transfer function as follows,

$$g(m\Delta\xi, n\Delta\eta) = e^{ikz} \mathbf{DFT}^{-1}\left\{\mathbf{DFT}[O(x,y)]\exp\left\{-i\pi\lambda z(s^2\Delta u^2 + t^2\Delta v^2)\right\}\right\}. \tag{31}$$

The sampling criterion can be interpreted from the analysis of a local frequency of the function $H(u,v)$ with a phase term, $\phi(u,v;z) = \pi\lambda z(u^2 + v^2)$. The maximum frequency $f_{u,\max}$ of the plane in the *u*-coordinate is given by

$$f_{u,\max} = \frac{1}{2\pi}\left|\frac{\partial\phi}{\partial u}\right|_{\max} = \lambda z|u|_{\max}. \tag{32}$$

To avoid an aliasing error, the sampling intervals $\Delta u$ should be satisfied with the condition, $\Delta u^{-1} \geq 2|f_{u,\max}|$ [15]. From this, we find that the sampling interval is not largely restricted by a short distance *z* other than the hologram synthesized from the conventional Fresnel transform in Eq. (5). However, the sampling rate is rather obstructed at a larger distance. In the Fresnel diffraction regime, the transfer function is identical with that of the angular spectrum method, where this aliasing effect has been studied in detail [28].

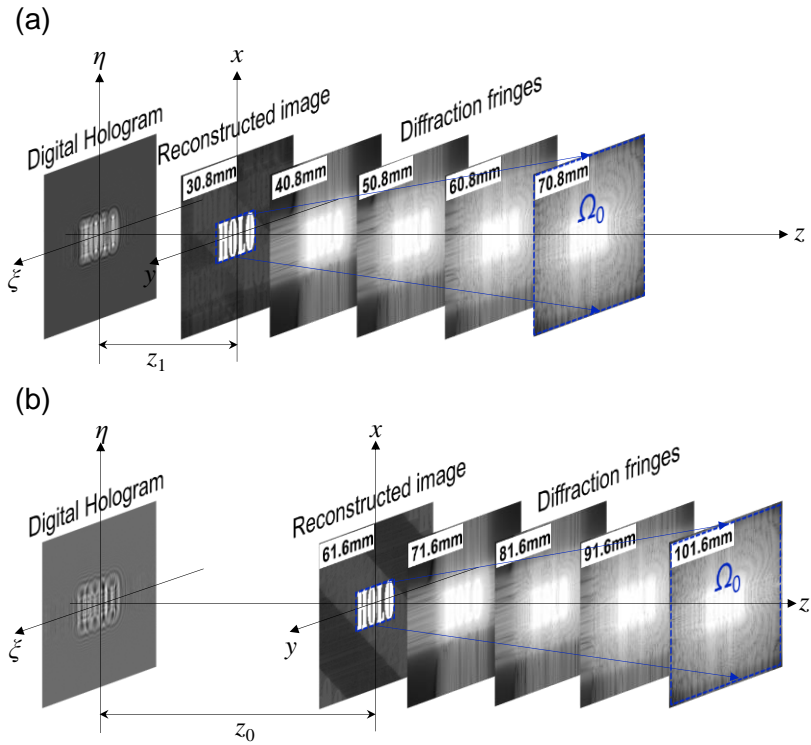

Fig. 9. Numerical results of the viewing-angle change in the reconstructed image from the hologram made at (a) a 30.8-mm distance and (b) 61.6-mm distance by using the convolution method. The reconstructed images located at different distances has the same field-size.

Figure 9 displays the numerical analysis of the AFOV for the reconstructed image from the hologram made by the convolution approach. The reconstructed image size located at different distances has the same value because of an identical pixel size of the hologram and object image. The hologram without an aliasing error is well obtained even at a short distance between the hologram and object. The specifications of the hologram synthesis are the same as those of the previous Fresnel transformation. The pixel size of the hologram is fixed to be 8 μm, and thus, all the images have 8-μm pixel. The diffraction fringes propagated from the hologram made at a 30.8-mm distance $z_1$ is illustrated in Fig. 9(a). The digital hologram consists of 512×512 pixels. The total field-size of all the diffraction fringes maintains to be 4096 μm, while the active diffraction fringe with respect to the letter image spreads out with increasing a reconstruction distance. The obtained viewing-angle from the increasing rate of the diffraction fringe is about to be 3.53°. Figure 9(b) is the simulation results for the hologram synthesized at a 61.6-mm distance $z_0$. Although the image is reconstructed at far away from the hologram plane, the active fringe diffracts at a similar rate with that in Fig. 9(a) where the viewing-angle is appeared to be 3.54°.

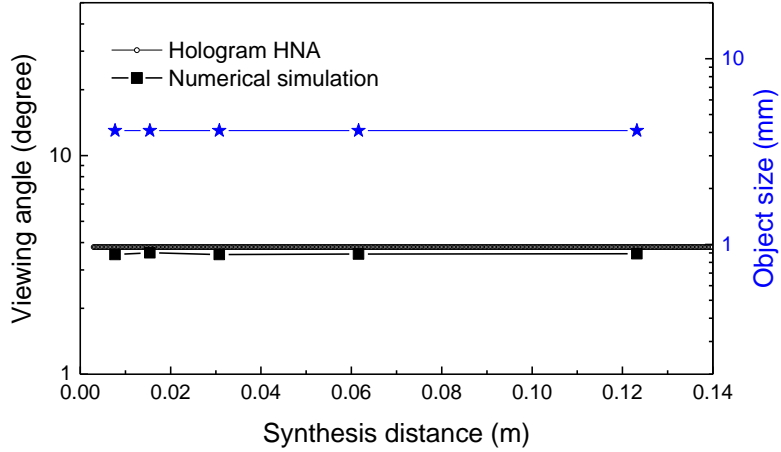

Fig. 10. Viewing-angle change in the reconstructed image for the digital holograms synthesized by using the convolution method. The object size does not vary irrespective of a synthesis distance.

Figure 10 shows the change of the viewing-angle in the holographic images as a function of a synthesis distance. All the angle values are similar irrespective of a synthesis-distance variation. Here, the resolution limit $R_{Abbe}$ has a constant value of 8 μm for all the images, which corresponds to the description in Fig. 4(b). The viewing-angle $\Omega$ is subject to Eq. (26). The value $\Omega$ of 3.81° is close to the measurement value. The schematic diagram of the numerical aperture for the digital hologram made by the convolution method is displayed in Fig. 11. The angle is not directly estimated from the lateral size of the digital hologram, unlike the result for the hologram made by the Fresnel transform method. As shown in Fig. 11(a), the hologram fringe does not fully occupy all the area of the digital hologram at a close distance from the object, where a low-pass filtering takes place due to a pixel pitch. In the hologram synthesis from the Fresnel transformation method, even a point source fills its total area except for undersampling of hologram fringe due to a finite pixel pitch. The convolutional method uses a double Fourier transform. In the hologram plane, the diffraction extent of 8-μm object resolution is conserved through an intermediate state of Fourier domain:

$$\Delta \xi = \frac{1}{N\Delta u} = \Delta x . \tag{33}$$

The size of the HNA is defined by a diffraction scope propagating from the object image. On the other hand, at a further distance in Fig. 11(c) the digital hologram will capture a partial diffractive wave. Nevertheless, the viewing-angle of a holographic image maintains constant, which is resulted from that the numerical reconstruction by the convolution method keeps a hologram resolution of 8 μm. We also observed this property in the holograms synthesized from the angular spectrum method, as not displayed here.

From above result, we note that in digital holography, the convolution method can numerically reconstruct a holographic image with a resolution of hologram pixel pitch even by using digital hologram occupied a part of the diffractive wave. In an optical system of hologram acquisition, since the object is analog, the diffraction ability from object resolution is not limited, and thus, digital hologram will be captured adaptively in accordance with the pixel specification of digital sensor. Considering its image reconstruction by using the conventional Fresnel transform, we can suppose that the image resolution will decrease with a reconstructed distance, which is another expression that as previously explained in Section 2, the sufficient aperture size is not secured due to a finite sensor size. We observed that this type of image resolution reduces at a further distance. However, the numerical convolutional method using a double Fourier transform overcomes this weakness of a resolution decrement.

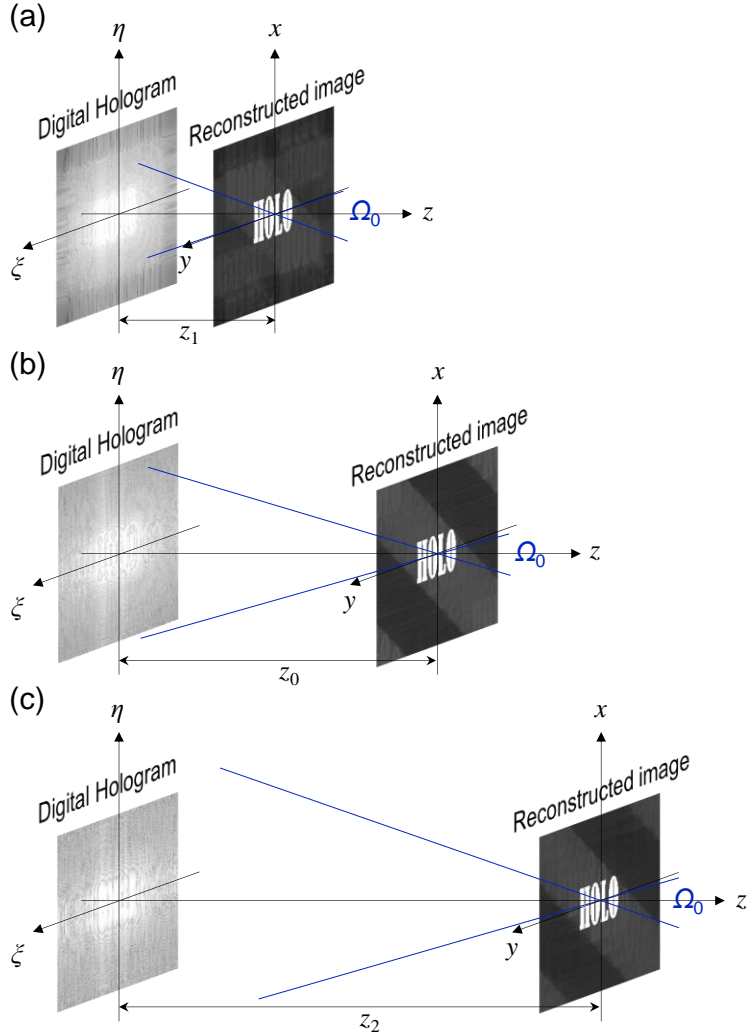

Fig. 11. Numerical aperture angle of the digital holograms synthesized at (a) 30.8-mm, (b) 61.6-mm, and (c) 123.2-mm distances using the convolutional approach. Digital holograms are displayed with a logarithmic scale.

## 4. Optical experiments for observing AFOV dependence on hologram numerical aperture

We carry out optical experiments to verify our analysis of the AFOV of a holographic image dependent on numerical aperture of a digital hologram. Figure 12(a) depicts the configuration of the object and digital hologram in a hologram synthesis. The hologram is synthesized by using the Fresnel transform method where based on Eq. (19), the object field size increases with a synthesis distance. To investigate the perspective view of 3-dimensional object, two letter objects placed at a different distance are vertically stacked on the coaxial $x$-axis. As analyzed in Section 2, the object located at $z_0$ has the same viewing-angle as the diffraction angle of a pixel pitch. If two objects are closer, the side-view of two objects is invisible in the reconstruction process, because as illustrated in Fig. 5, the diffractive fringe of the former letter object image propagates at a diffraction angle, and thereafter covers almost the same area of the latter object image at $z_0$-distance. When the object location is larger than $z_0$-distance, we cannot see the side-view of object image in this situation. However, at a closer distance, we find that the viewing-angle of a holographic image is higher than the diffraction angle, and thus, the lateral view of two letter object images would be visible.

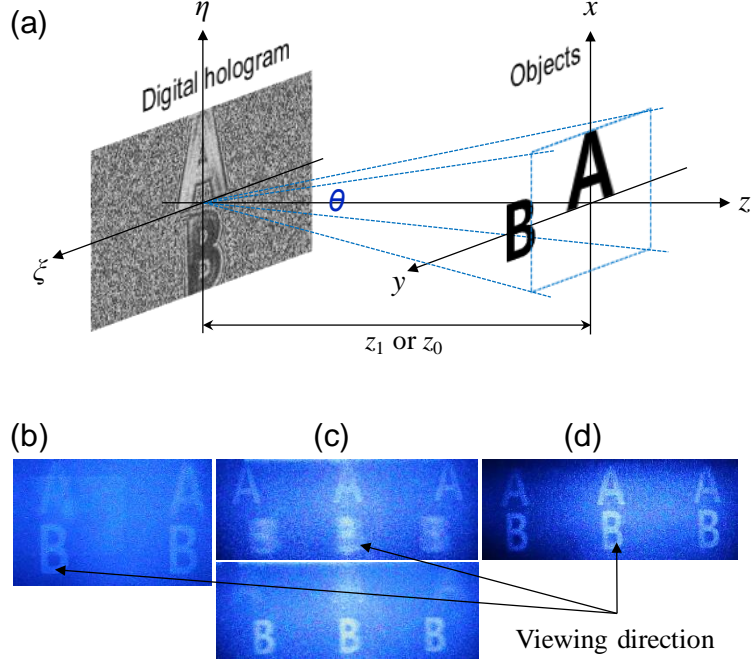

Fig. 12. Optical experiments for the AFOV of a holographic image dependent on the hologram numerical aperture. (a) Configuration of objects and digital hologram. Reconstructed images from digital hologram synthesized (b) at $z_0$-distance. Reconstructed images from digital hologram synthesized at $z_1$-distance for two letter objects placed (c) at the different distance and (d) at the same distance.

Figure 12(b) shows the optically reconstructed image for the digital hologram synthesized at $z_0$-distance. We use the phase SLM (Holoeye PLUTO) with 1920×1080 pixels of 8-μm pixel pitch. To obtain the adequate phase hologram, the digital hologram is synthesized through Gerchberg-Saxton iterative algorithm [27,29]. The $z_0$-distance of Eq. (28) for lateral direction is calculated to be 259.8 mm and two objects are spaced 30 mm apart. The collimated plane wave from the blue laser with 473 nm illuminates the reflective panel through beam splitter. The high-order images are captured within the lens aperture of DSLR camera, where the removal of noises is not considered because they do not affect the viewing-angle analysis. The images are captured at a particular viewing direction, and thus, the images adjacent to the central image captured at a viewing direction display their perspective views at a different viewing direction. Refocus phenomenon of two letters in accordance with a depth is clearly confirmed. Here, the side-view of adjacent image cannot be observable, which is resulted from the fact that as described above, the viewing-angle is not higher than the diffraction angle of 3.8° of a pixel pitch.

Figure 12(c) displays the reconstructed image for the digital hologram synthesized at $z_1$-distance of 129.9 mm. According to our numerical analysis, the reconstructed image has 4-μm resolution and the viewing-angle of 7.6°. We can observe the lateral view of two objects in the adjacent images, which indicates that the AFOV is higher than diffraction range of a pixel pitch.

We also confirm the variation of perspective view of central image by moving the viewing-direction to the adjacent image. Below figure shows the reconstructed image focused on the former letter. When two letter objects are placed at the same location, the perspective view does not appear, as displayed in Fig. 12(d). From the maximum perspective view of the reconstructed image, the viewing-angle is estimated to be 7.8°. The experimental result supports our analysis that the viewing-angle of a holographic image is fundamentally determined by numerical aperture of a digital hologram. Only the diffraction angle of a pixel pitch obstructs to secure an observable viewing window without the superposition of aliased images. It is very interesting that the high viewing-angle can be realized by blocking the adjacent images as long as the image is nearly small, which looks like a similar behavior to the trade-off relation of Eq. (1). However, in this case, if the diffraction angle of a pixel pitch can be expanded, the sufficient viewing window would be formed, which will be discussed in subsequent Section.

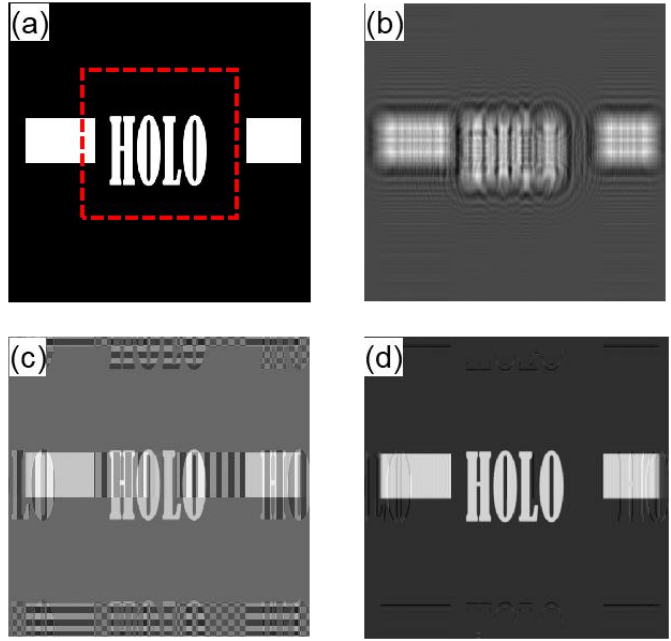

Fig. 13. Simulation result for securing the AFOV of the holographic image with an enlarged size. (a) Object and (b) its hologram synthesized by the Riemann integral in the Rayleigh-Sommerfeld diffraction formula. The reconstructed images (c) from original hologram and (d) from upsampled hologram fringe.

## 5. Viewing-angle expansion of holographic image and its discussion

The AFOV in the hologram made using the Fresnel transform method increases with decreasing a synthesis distance, where as previously described in Fig. 1, it is inevitable that the object size decreases in a high-HNA hologram synthesis to avoid an aliasing effect. This description can be naturally extended to the Rayleigh-Sommerfeld diffraction scheme. The sampling rate $f_s$ of the object field follows the Nyquist criterion, $f_s \geq 2 f_{x,\max}$. The sampling pixel size $\Delta x$ should satisfy the following condition [30]:

$$\frac{1}{\Delta x} \geq 2 \frac{|\xi - x|_{\max}}{\lambda \sqrt{(\xi - x)^2 + z^2}}. \tag{34}$$

The sampling pixel value depends on the calculated field-size $|\xi - x|_{max}$ as well as a synthesis distance. As the calculated field size increases, the finer pixel sampling is required.

Figure 13 is the simulation result for securing the AFOV of a holographic image with an enlarged size. We consider the digital hologram of 256×256 size synthesized at the $z_1$ distance of 15.4 mm in Fig. 1. The red box in Fig. 13(a) indicates the letter object with 256×256 pixel of a 4-μm resolution. The rectangular object is added to the outside of the letter object, and the whole object with 512×512 pixel has the same size as 2048-μm of the hologram. Based on Eq. (34), the hologram can be calculated from the object with an enlarged size through its upsampling process. The digital hologram with no aliasing error is obtained through a two-fold upsampling process of the object, which is calculated from the Riemann integral, as illustrated in Fig. 13(b). This upsampling process is different from that in Eq. (29), which means that the object itself is sampled to the finer pixel. We find that the lateral size of the object is not critical compared to the synthesis distance, where a no-aliasing hologram is achieved even without the upsampling operation. In this upsampling case, we also notice that although the object resolution increases up to 2-μm, the HNA of the hologram is not affected. The Rayleigh-Sommerfeld diffraction formula is the extension of the Fresnel approximation. The diffraction extent in the hologram plane doubles where half of diffractive wave becomes a digital hologram.

The display process of the holographic scene is also subject to the criterion of Eq. (34). Figure 13(c) illustrates the reconstructed image without its upsampling of the hologram fringe. The aliasing noise images are overlapped with the original image. These aliasing images are generated from the high-order diffraction beams due to its pixel pitch of a modulator, where the diffraction zone is a half of the field-view. When the digital hologram is upsampled from 256×256 pixel of 8-μm resolution to 512×512 pixel of 4-μm resolution, the object image with an extended field is well retrieved in Fig. 13(d), but the aliasing error of high-order terms is not completely removed. The viewing-angle calculated from diffraction fringes is appeared to be 7.3°, which is double the value for 8-μm pixel diffraction.

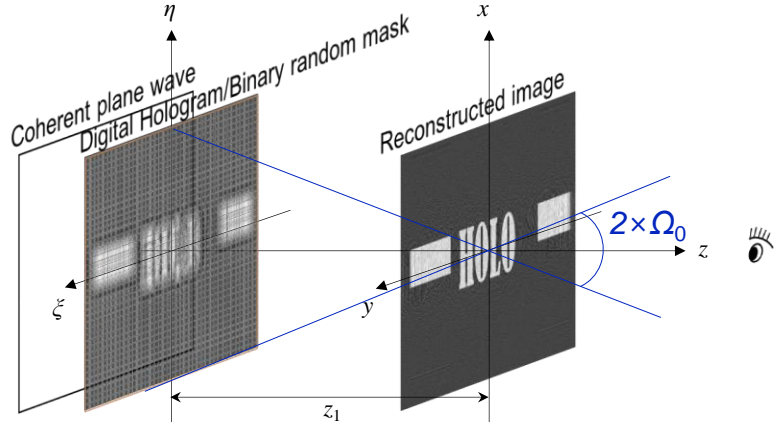

Fig. 14. Simulation result for holographic display with an enhanced viewing-angle by using a binary random mask.

Figure 14 shows simulation results for holographic display with an enhanced viewing-angle. The binary random mask with 512×512 pixel of 4-μm resolution is used to eliminate the aliasing noise images. Although the opening ratio of mask is put to be 90%, we obtain a reconstructed image largely removed noise images. The random sampling deteriorates a periodicity of the pixel structure, and thus prohibits the formation of high-order diffraction patterns in the Fourier space. Therefore, we find that only the adding its randomness to the upsampled hologram effectively removes the aliasing images.

This shows the possibility of the viewing-angle enhancement of the holographic image with a present modulator. In a real system, the binary random mask could be manufactured by coating the black-matrix grating on a transparent substrate. The high-HNA digital hologram can be synthesized within the specification of a present modulator, where the high-order aliasing images are appeared at the outside of the diffraction zone. These aliasing images could be effectively eliminated by upsampling the digital hologram through a binary random mask. Another way to remove the aliasing images is to design the spatial modulator itself with randomly distributed pixels. These approaches could be a useful tool to develop the wide viewing-angle holographic display [31].

## 6. Conclusion

We elucidate that the AFOV of holographic images is determined from the HNA rather than a pixel pitch. In other words, the resolving power of the digital hologram becomes a key factor for an ability of the angular view. The numerical simulation and optical experiments for various types of holograms prove that the viewing-angle strongly depends on the HNA, where the hologram with a large numerical aperture reconstructs the image with a high viewing-angle. We demonstrate that the holographic display with a wide viewing-angle could be realized by using a high-HNA hologram and removing the aliasing noise images. The high-HNA hologram can be synthesized using the object field beyond the diffraction zone of a pixel pitch, where the high-order aliasing images appeared outside of the diffraction zone could be effectively eliminated by upsampling the digital hologram through a binary random mask.

## Acknowledgments

This work was partially supported by Institute for Information & Communications Technology Promotion (IITP) grant funded by the Korea government (MSIP) (2017-0-00049)